\documentclass[twocolumn,showpacs,preprintnumbers,
amsmath,amssymb,aps,prd,nofootinbib,superscriptaddress,
eqsecnum]{revtex4}
\usepackage{graphicx}

\makeatletter
\renewcommand{\p@subsection}{}
\makeatother

\newcommand{\Slash}[1]{\ooalign{\hfil/\hfil\crcr$#1$}}

\begin{document}

\title{Thermodynamics of dense hadronic matter in a parity doublet model}

\author{Chihiro Sasaki}
\affiliation{%
Frankfurt Institute for Advanced Studies,
D-60438 Frankfurt am Main,
Germany
}
\author{Igor Mishustin}
\affiliation{%
Frankfurt Institute for Advanced Studies,
D-60438 Frankfurt am Main,
Germany
}
\affiliation{%
Kurchatov Institute, Russian Research Center,
Moscow 123182, Russia
}

\date{\today}

\begin{abstract}
We study thermodynamics of nuclear matter in a two-flavored
parity doublet model within the mean field approximation.
Parameters of the model are chosen to reproduce correctly
the properties of the nuclear ground state.
The model predicts two phase transitions in nuclear matter,
a liquid-gas phase transition at normal nuclear density 
and a chiral transition at higher density.
At finite temperature the pion decay constant exhibits a 
considerable reduction at intermediate values of chemical potential, which is 
traced back to the presence of the liquid-gas transition,
and approaches zero at higher chemical potential associated with
the chiral symmetry restoration.
A ``transition'' from meson-rich to baryon-rich matter is also discussed.
\end{abstract}


\maketitle

%%%%%%%%%%%%%%%%%%%%%%%%%%%%%%%%%%%%%%%%%%%%%%%%%%%%%
\section{Introduction}
\label{sec:int}
%%%%%%%%%%%%%%%%%%%%%%%%%%%%%%%%%%%%%%%%%%%%%%%%%%%%

Model studies of dense quark matter have suggested a rich phase structure 
of QCD at temperatures and quark chemical potentials of order 
$\Lambda_{\rm QCD}$. Our knowledge on the phase structure however remains 
limited and the description of strongly interacting matter does not 
reach a consensus yet~\cite{qmproc}. 
In particular, properties of baryons near the chiral symmetry restoration are
poorly understood. 
We believe that the realistic modeling of dense baryonic matter must take into
account the existence of the nuclear matter saturation point, i.e. the bound
state at baryon density $\rho_0 = 0.16$ fm$^{-3}$, like in Walecka type
models~\cite{walecka}.
Several chiral models with pure hadronic degrees of freedom~\cite{other,nnjl}
have been constructed in such a way that the nuclear matter has the true 
ground state. Other models consider a nucleon as
a dynamical bound-state of a diquark and a quark~\cite{bentz}.

In chiral models it is mostly {\it assumed} that a constituent quark
transforms under chiral rotations as a bare quark. According to this naive 
assignment, dynamical chiral symmetry breaking entirely generates a baryon 
mass which thus vanishes at the restoration.
According to an alternative, mirror assignment~\cite{dk,mirror},
dynamical chiral symmetry breaking generates a mass difference between parity 
partners and the chiral symmetry restoration does not necessarily dictate the 
chiral partners to be massless. Mirror baryons embedded in linear and non-linear 
chiral Lagrangians have been applied to study their phenomenology 
in vacuum~\cite{dk,mirror,lsma1}, nuclear matter~\cite{hp,pdm} 
and neutron starts~\cite{astro}, although it remains an open question which 
scenario is preferred by nature. Identifying the true parity partner of 
a nucleon is also an issue. In the mirror models $N(1535)$ is usually taken to 
be the negative parity state. This choice however fails to reproduce the decay 
width to a nucleon and $\eta$. This might indicate another negative parity 
state lighter than the $N(1535)$~\cite{pdm}, which has not been observed so far.

In this paper we apply the parity doublet model to a hot and dense hadronic 
matter and study the phase structure of a chiral phase transition as well as 
a liquid-gas transition of nuclear matter. The main interest is the phase
boundary of the chiral symmetry restoration on top of the nuclear matter
ground state. A light negative parity partner will be considered as a 
phenomenological option and we show how the phase boundary lines are affected. 
The phase transitions in large number of colors are also discussed.

%%%%%%%%%%%%%%%%%%%%%%%%%%%%%%%%%%%%%%%%%%%%%%%%%%%%%
\section{Parity doublet model}
\label{sec:model}
%%%%%%%%%%%%%%%%%%%%%%%%%%%%%%%%%%%%%%%%%%%%%%%%%%%%

The negative-parity nucleon field is introduced such that
it transforms in the opposite way as the positive parity
nucleon under SU(2)$_L \times$ SU(2)$_R$~\cite{dk}:
\begin{eqnarray}
&&
\psi_{+L} \to L\,\psi_{+L}\,,
\quad
\psi_{+R} \to R\,\psi_{+R}\,,
\nonumber\\
&&
\psi_{-L} \to R\,\psi_{-L}\,,
\quad
\psi_{-R} \to L\,\psi_{-R}\,,
\end{eqnarray}
with $L \in$ SU(2)$_L$ and $R \in$ SU(2)$_R$, and
this allows a chirally invariant mass term.
The Lagrangian of the mirror baryons is given by
\begin{eqnarray}
{\mathcal L}
&=& \bar{\psi}_+ i\Slash{\partial}\psi_+
{}+ \bar{\psi}_- i\Slash{\partial}\psi_-
{}+ m_0\left( \bar{\psi}_- \gamma_5 \psi_+ 
{}- \bar{\psi}_+ \gamma_5 \psi_- \right)
\nonumber\\
&&
{}+ a \bar{\psi}_+ 
\left( \sigma + i\gamma_5\vec{\tau}\cdot\vec{\pi}\right)\psi_+
{}+ b \bar{\psi}_- 
\left( \sigma - i\gamma_5\vec{\tau}\cdot\vec{\pi}\right)\psi_-
\nonumber\\
&&
{}- g_\omega\bar{\psi}_+\gamma_\mu\omega^\mu\psi_+
{}- g_\omega\bar{\psi}_-\gamma_\mu\omega^\mu\psi_-
{}+ {\mathcal L}_M\,,
\label{lag}
\end{eqnarray}
where $a, b$ and $g_\omega$ are the coupling constants of
the mesons to the baryons. 
The mesonic part follows the standard linear sigma model Lagrangian,
\begin{eqnarray}
{\mathcal L}_M
&=& \frac{1}{2}\partial_\mu\sigma\cdot\partial^\mu\sigma
{}+ \frac{1}{2}\partial_\mu\vec{\pi}\cdot\partial^\mu\vec{\pi}
{}- \frac{1}{4}F_{\mu\nu}F^{\mu\nu}
\nonumber\\
&&
{}+ \frac{1}{2}m_\omega^2\omega_\mu\omega^\mu
{}+ g_4^4\left( \omega_\mu\omega^\mu \right)^2
\nonumber\\
&&
{}+ \frac{1}{2}\bar{\mu}^2\left( \sigma^2 + \vec{\pi}^2 \right)
{}- \frac{1}{4}\lambda\left( \sigma^2 + \vec{\pi}^2 \right)^2
{}+ \epsilon\sigma\,,
\end{eqnarray}
with the field strength tensor of the omega meson, 
$F_{\mu\nu} = \partial_\mu\omega_\nu - \partial_\nu\omega_\mu$.
The parameters $\lambda, \bar{\mu}$ and $\epsilon$ are related to
the sigma and pion masses and the pion decay constant in vacuum as
\begin{eqnarray}
\lambda &=& \frac{m_\sigma^2 - m_\pi^2}{2\sigma_0^2}\,,
\nonumber\\
\bar{\mu}^2 &=& \frac{m_\sigma^2 - 3m_\pi^2}{2}\,,
\nonumber\\
\epsilon &=& m_\pi^2 f_\pi\,,
\end{eqnarray}
with $m_\pi = 138$ MeV, $f_\pi = 93$ MeV and the vacuum expectation
value of the sigma field $\sigma_0 = f_\pi$. The vacuum mass of
the omega field is $m_\omega = 783$ MeV. Other mesonic parameters, 
$m_\sigma$ and $g_4$, as well as the chiral invariant mass $m_0$ will
be fixed to reproduce the known properties of nuclear matter.

The mass term of (\ref{lag}) is diagonalized by the mass eigenstates
of the parity doubled nucleons, $N_+$ and $N_-$ via
\begin{equation}
\begin{pmatrix}
N_+
\\
N_-
\end{pmatrix}
=
\frac{1}{\sqrt{2\cosh\delta}}
\begin{pmatrix}
e^{\delta/2} & \gamma_5 e^{-\delta/2}
\\
\gamma_5 e^{-\delta/2} & -e^{\delta/2}
\end{pmatrix}
\begin{pmatrix}
\psi_+
\\
\psi_-
\end{pmatrix}\,,
\end{equation}
with $\sinh\delta = -(a+b)\sigma/2m_0$.
The masses of $N_\pm$ are thus given by
\begin{equation}
m_{N\pm} = \frac{1}{2}\left[ 
\sqrt{(a+b)^2\sigma^2 + 4m_0^2} \mp (a-b)\sigma
\right]\,.
\label{bmass}
\end{equation}
When the chiral symmetry is restored ($\sigma_0 = 0$),
the two states become degenerate, $m_{N+} = m_{N-} = m_0$.
In contrast, in the naive transformation under chiral symmetry,
i.e. $\psi_{\pm L} \to L\,\psi_{\pm L}$ and $\psi_{\pm R} \to R\,\psi_{\pm R}$,
the nucleon mass is generated solely by
the spontaneous chiral symmetry breaking and thus vanishes
with its restoration.

Applying a mean field approximation~\footnote{
  Rotational invariance requires $\langle \vec{\omega} \rangle = 0$
  and only the time-component $\langle \omega_0 \rangle$ does not vanish.
}, 
the thermodynamic potential is
\begin{eqnarray}
\Omega 
&=& \sum_{i=\pm}\gamma_i \int\frac{d^3\vec{p}}{(2\pi)^3}
T \left[ \ln\left( 1 - n_f(T,\mu^\ast;m_i) \right)
\right.
\nonumber\\
&&
\left.
{}+ \ln\left( 1 - \bar{n}_f(T,\mu^\ast;m_i) \right)
\right]
\nonumber\\
&&
{}- \frac{1}{2}m_\omega^2\omega_0^2 - g_4^4\omega_0^4
{}- \frac{1}{2}\bar{\mu}^2\sigma^2 + \frac{1}{4}\lambda\sigma^4
{}- \epsilon\sigma\,,
\label{omega}
\end{eqnarray}
where $\gamma_i$ denotes the spin-isospin degeneracy factor
of a nucleon and $n_f$ ($\bar{n}_f$) is the Fermi-Dirac distribution 
function for a fermion (an anti-fermion) defined by
\begin{eqnarray}
n_f 
&=& \frac{1}{1 + e^{(E_i - \mu^\ast)/T}}\,,
\nonumber\\
\bar{n}_f 
&=& \frac{1}{1 + e^{(E_i + \mu^\ast)/T}}\,,
\end{eqnarray}
with the energy $E_i = \sqrt{\vec{p}^2 + m_i^2}$ and
the effective chemical potential $\mu^\ast = \mu - g_\omega\omega_0$.
Imposing the stationary condition, $\partial\Omega/\partial\sigma
= \partial\Omega/\partial\omega_0 = 0$, one obtains the coupled
field equations:
\begin{eqnarray}
&&
\sum_{i=\pm}\gamma_i \int\frac{d^3\vec{p}}{(2\pi)^3}
\frac{m_i}{E_i}\frac{\partial m_i}{\partial\sigma}
\left[ n_f + \bar{n}_f \right]
{}- \bar{\mu}^2\sigma + \lambda\sigma^3 - \epsilon = 0\,,
\nonumber\\
&&
g_\omega \sum_{i=\pm}\gamma_i \int\frac{d^3\vec{p}}{(2\pi)^3}
\left[ n_f - \bar{n}_f \right]
{}- m_\omega^2\omega_0 - 4g_4^4\omega_0^3 = 0\,.
\end{eqnarray}
The net baryon number density is given by
\begin{equation}
\rho 
= \sum_{i=\pm}\gamma_i
\int\frac{d^3\vec{p}}{(2\pi)^3}
\left[ n_f - \bar{n}_f \right]\,.
\end{equation}

The model parameters are fixed to reproduce the nuclear matter saturation
properties at $T = 0$:
\begin{eqnarray}
&&
E/A(\rho_0) - m_{N+} = - 16\,\mbox{MeV}\,,
\nonumber\\
&&
\rho_0 = 0.16\,\mbox{fm}^{-3}\,.
\label{binding}
\end{eqnarray}
Originally, 
the negative parity state has been chosen to be $N(1535)$~\cite{dk,mirror}.
However this choice is incompatible with the width of decay into $\eta N$.
Also, the mass difference between $N(1535)$ and $N_+$ is about
$560$ MeV, that is well above the QCD scale $\Lambda_{\rm QCD} \sim 200$ MeV.
The spontaneous chiral symmetry breaking might not play much role on the 
physics of such excited states.
Another candidate, which is lighter and has a stronger coupling to the pion,
is considered to be a phenomenological option~\cite{pdm,astro}.
We will thus consider two sets of model parameters as given in~\cite{pdm} 
(see Table~\ref{para}):
%%%%%%%%%%%%%%%%%%%%%%%%%%%%%%%%%%%%%%%%%
\begin{table}
\begin{center}
\begin{tabular*}{6cm}{@{\extracolsep{\fill}}cccccc}
\hline
{} & set A & set B \\
\hline
$m_{N-}\,\mbox{[MeV]}$ & $1500$ & $1200$ \\
$m_0\,\mbox{[MeV]}$    & $790$  & $790$  \\
$m_\sigma\,\mbox{[MeV]}$ & $370.63$ & $318.56$ \\
$g_\omega$           & $6.79$ & $6.08$ \\
$a$                  & $13.00$ & $9.16$ \\
$b$                  & $6.97$ & $6.35$ \\
$\bar{\mu}\,\mbox{[MeV]}$ & $199.26$ & $147.50$ \\
$\lambda$            & $6.82$ & $4.75$ \\
\hline
\end{tabular*}
\end{center}
\caption{
Set of parameters with $g_4=0$~\cite{pdm}. 
The vacuum nucleon mass is $m_{N+}=935$ MeV.
}
\label{para}
\end{table}
%%%%%%%%%%%%%%%%%%%%%%%%%%%%%%%%%%%%%%%%%%
With these sets of parameters the incompressibility defined by
\begin{equation}
K = 9\rho_0^2\frac{\partial^2(E/A)}{\partial\rho^2}\Big{|}_{\rho_0}
{}= 9\frac{\partial P}{\partial\rho}\Big{|}_{\rho_0}
{}= 9\rho_0\frac{\partial\mu_B}{\partial\rho}\Big{|}_{\rho_0}\,,
\label{incompress}
\end{equation}
is $430$-$510$ MeV. This is slightly higher but still compatible to the 
suggested range, 
$200$-$400$ MeV, from the analysis of giant mono-pole resonances~\cite{compress}.
Note that finite $g_4$ decreases the incompressibility, 
e.g. $K = 440$ MeV ($m_{N-}=1500$ MeV) and $K = 370$ MeV ($m_{N-}=1200$ MeV)
for $g_4=3.8$~\cite{pdm}. This however does not change much either
the (pseudo)critical density or the order of phase transitions.
For simplicity of our calculations, we will therefore take $g_4=0$.

%%%%%%%%%%%%%%%%%%%%%%%%%%%%%%%%%%%%%%%%%%%%%%%%%%%%%
\section{Thermodynamics of the model}
\label{sec:thermo}
%%%%%%%%%%%%%%%%%%%%%%%%%%%%%%%%%%%%%%%%%%%%%%%%%%%%

\begin{figure}
\begin{center}
\includegraphics[width=8cm]{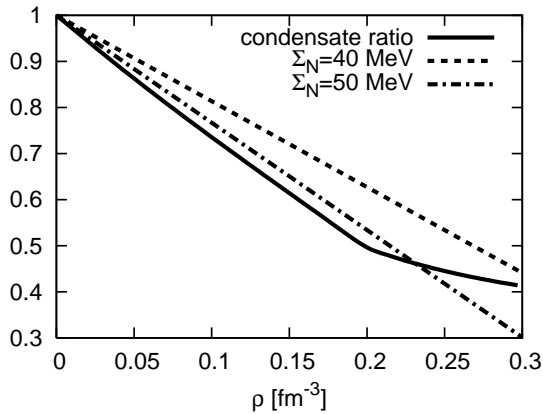}
\caption{
The ratio of the $\bar{q}q$ condensate in medium
to its vacuum value. The calculation was done at $T=0$
with the parameter set A. 
For set B there is no significant difference from the above result.
}
\label{cond}
\end{center}
\end{figure}

The density dependence of the quark condensate $\langle \bar{q}q \rangle$
can be extracted using the Feynman-Hellmann theorem and it is model
independent to the leading order in nucleon density~\cite{qqbar},
\begin{equation}
\frac{\langle \bar{q}q \rangle_\rho}{\langle \bar{q}q \rangle_{\rm vac}}
= 1 - \frac{\Sigma_N}{m_\pi^2 f_\pi^2}\rho\,,
\label{qqbar}
\end{equation}
with the nucleon sigma term $\Sigma_N = 45 \pm 8$ MeV~\cite{sigma}.
Fig.~\ref{cond} shows a comparison of the low-energy theorem (\ref{qqbar})
with the in-medium quark condensate calculated in the parity doublet model.
%%%%%%%%%%%%%%%%%%%%%%%%%%%%%%%%%%%%%%%%%%%%%
%%%%%%%%%%%%%%%%%%%%%%%%%%%%%%%%%%%%%%%%%%%
The model describes the correct linearity at low densities up to around
the saturation density $\rho_0$. The value of the condensate, however, 
decreases somewhat stronger than Eq.~(\ref{qqbar}).
This can be considered as an artifact of our simple model and taking other
parameters (with non-vanishing $g_4$) or more systematic treatment would make 
a better agreement with the empirical value of $\Sigma_N$.
In-medium chiral condensate has been explored in chiral perturbation theory 
constrained by the nuclear matter properties and shown to have a strong pion
mass dependence, and a substantial deviation from the linear density approximation
was found already at $\rho \sim 0.2$ fm$^{-3}$ where the ratio is 
$\sim 0.7$~\cite{chpt}. In that framework 
the two-pion exchange correlations with virtual $\Delta(1232)$ resonances are 
responsible for stabilizing the quark condensate, which are not incorporated 
in the current mean field model.

It is remarkable that the parity doublet model at $T=0$ 
describes a liquid-gas transition of nuclear matter and a first-order 
(set A) or crossover type (set B) of chiral symmetry restoration~\cite{pdm}.
At an intermediate temperature, above these phase transitions, the order
parameter typically shows a double-step structure as shown in Fig.~\ref{fpi}:
%%%%%%%%%%%%%%%%%%%%%%%%%%%%%%%%%%%%%%%%%%%%%
\begin{figure}
\begin{center}
\includegraphics[width=8cm]{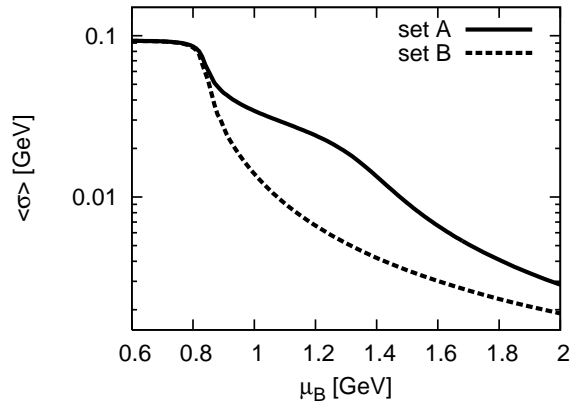}
\caption{
The expectation value of the sigma field as a function of 
baryon chemical potential at $T=50$ MeV.
}
\label{fpi}
\end{center}
\end{figure}
%%%%%%%%%%%%%%%%%%%%%%%%%%%%%%%%%%%%%%%%%%%
The first drop in $\langle \sigma \rangle$ appears much below the chiral 
crossover and this is traced back to the remnant of the liquid-gas phase 
transition. The second drop in $\langle \sigma \rangle$ happens at a higher 
$\mu_B$ associated with 
the chiral symmetry restoration. The two-step structure is observed
only at rather low temperatures, and at high temperatures $\langle \sigma \rangle$
eventually experiences a single reduction due to the chiral crossover.
Note, as mentioned above, the first reduction will weaken when
improvements are included to reproduce Eq.~(\ref{qqbar}).

In the parity doublet model one can deal with in-medium mesons and 
baryons explicitly. Besides the baryon masses given in Eq.~(\ref{bmass}),
we define the effective meson masses in matter by
\begin{equation}
M_\sigma^2 = \frac{\partial^2\Omega}{\partial\sigma^2}\,,
\quad
M_{\vec{\pi}}^2 = \frac{\partial^2\Omega}{\partial\vec{\pi}^2}\,,
\quad
M_\omega^2 = -\frac{\partial^2\Omega}{\partial\omega_0^2}\,,
\label{mmass}
\end{equation}
where $\Omega$ is given in Eq.~(\ref{omega}).
The total meson number density is given by
\begin{equation}
\rho_M = \sum_{i=\sigma,\pi,\omega_0}\gamma_i
\int\frac{d^3\vec{p}}{(2\pi)^3}n_b(T;M_i)\,,
\end{equation}
where $\gamma_i$ is the corresponding mesonic degeneracy factors,
and $n_b$ is the Bose-Einstein distribution function,
\begin{equation}
n_b = \frac{1}{e^{E_i/T}-1}\,,
\end{equation}
with the corresponding particle energy $E_i = \sqrt{\vec{p}^2 + M_i^2}$.
For simplicity, here we consider only $\sigma, \pi$ and $\omega$ mesons,
although other mesons, especially $\rho$ with $\gamma_\rho=9$ may somewhat
contribute too.
The meson density depends on temperature and it is instructive to
compare it with the baryon density. We consider the ratio
%The particle number densities measure which particle species are
%thermodynamically populated. Therefore, we define the following ratio
of the meson to the baryon number density,
\begin{equation}
\frac{\rho_M}{\rho_B} = 1\,,
\label{mb}
\end{equation}
to determine a trajectory in the $T$-$\mu_B$ plane.
This trajectory separates a meson-dominated
region from a baryon-dominated one and characterizes a meson-to-baryon
``transition'' in dense matter. This is similar to the definition
adopted in Ref.~\cite{triple} where the statistical
model parameters $T$ and $\mu_B$ were extracted from the hadron yield ratios.
While the statistical
model successfully describes the particle abundances at chemical freeze-out,
it cannot handle the chiral phase transition. In the parity doublet model,
on the other hand, the nuclear matter ground state is built in and
the chiral symmetry restoration is also described. Thus, the ratio
(\ref{mb}) can be used to find the critical points on the phase diagram
%is expected to have a relevant density effect carried by 
%in-medium hadrons 
within a self-consistent chiral approach.

\begin{figure*}
\begin{center}
\includegraphics[width=8cm]{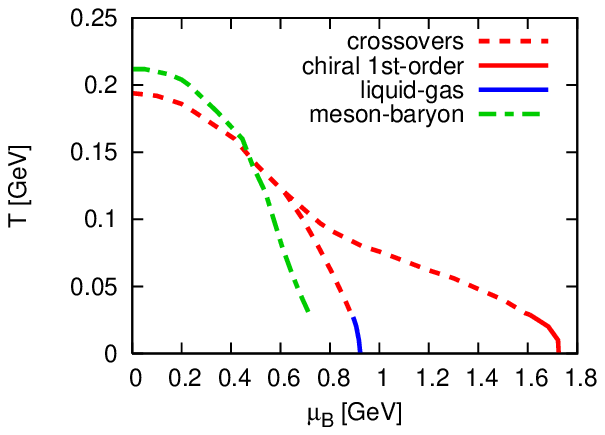}
\includegraphics[width=8cm]{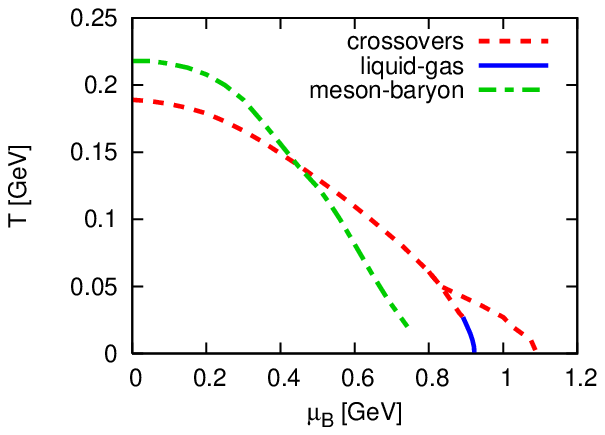}
\caption{
The phase diagram for the set A (left) and for the set B (right).
The crossovers correspond to the local/global maxima in the order 
parameter susceptibility.
}
\label{phase}
\end{center}
\end{figure*}

In Fig.~\ref{phase} we show the phase diagram of the parity doublet model
for the two parameter sets discussed above.
%%%%%%%%%%%%%%%%%%%%%%%%%%%%%%%%%%%%%%%%%%%%%
%%%%%%%%%%%%%%%%%%%%%%%%%%%%%%%%%%%%%%%%%%%
At zero temperature the system experiences a first-order liquid-gas
transition at $\mu_B = 923$ MeV and the baryon density shows a jump from
zero to a finite value $\rho \neq 0$. The order of chiral phase transition and its
location depend on the set of parameters, especially on mass of the negative 
parity state~\cite{pdm}. 
Roughly speaking this phase transition occurs when the baryon chemical potential
reaches the mass of the negative parity state, $\mu_B \sim m_{N-}$.
If we take the most frequently used value
%the candidate of the parity partner usually 
%taken with 
$m_{N-}=1500$ MeV, then in addition to the nuclear liquid-gas phase transition
we obtain a weak first-order chiral transition 
at $\rho \sim 10\,\rho_0$. 
With a lower mass $m_{N-}=1200$ MeV 
we get no true chiral phase transition but only a crossover at much lower density
$\rho \sim 3\,\rho_0$~\footnote{
  Finite $g_4 = 3.8$, which gives a slightly better fit for the 
  incompressibility $K$, does not change its order, but makes a first-order chiral 
  transition somewhat stronger~\cite{pdm}.
}.
The liquid-gas transition survives up to $T = 27$ MeV. Above this temperature
there is no sharp phase transition but the order parameter is still attracted 
by the critical point, corresponding to the first drop in $\langle \sigma \rangle$.
This makes an additional crossover line terminating at the liquid-gas 
critical point. Another crossover line corresponding to the chiral symmetry 
restoration follows 
the steepest descent of the second reduction in $\langle \sigma \rangle$. 
With increasing
temperature the two crossover lines become closer and finally merge.
The former crossover line, associated with a trace of the liquid-gas phase
transition, is not well determined from the susceptibility when
two bumps are getting closer to each other.

In contrast, the trajectory defined in Eq.~(\ref{mb}) is basically driven by
the density effect with the hadron masses being not far from their vacuum
values. The line is almost independent of the parameter set and goes rather
close to the liquid-gas transition line.
The chiral crossover and the meson-baryon transition lines intersect
at $(T,\mu_B) \sim (150,450)$ MeV. 
%One also sees that the line of
%Eq.~(\ref{mb}) and the crossover attached to the liquid-gas transition appear
%near. 
The parity doublet model thus describes 3 domains: a chirally broken
phase with mesons thermodynamically dominating, another chirally broken phase 
where baryons are more dominant and the chirally restored phase,
which can be identified with quarkyonic matter~\cite{quarkyonic,mrs}.
It is worthy to note that this point is fairly
close to the estimated triple point at which hadronic matter, quarkyonic matter
and quark-gluon plasma may coexist~\cite{triple}.

%%%%%%%%%%%%%%%%%%%%%%%%%%%%%%%%%%%%%%%%%%%%%%%%%%%%%
\section{Chiral crossover in nucleonic NJL model}
\label{sec:other}
%%%%%%%%%%%%%%%%%%%%%%%%%%%%%%%%%%%%%%%%%%%%%%%%%%%%

There exist other chiral models which have attempted to describe 
nuclear matter ground state~\cite{other,nnjl}. The vector-type interactions 
at finite density, isoscalar vector and a mixed scalar-vector ones, 
play a significant role to reproduce correctly the nuclear saturation point.
Fermions in those works were introduced in the naive assignment and 
the nucleon thus becomes massless in the chiral limit when chiral symmetry 
is restored.
In this section we give a brief discussion about the thermodynamics at finite
temperatures of a nucleonic Nambu--Jona-Lasinio model (NJL)~\cite{nnjl}
and compare to the result in the parity doublet model.

At $\mu_B = 0$ the vector interaction does not affect the gap equation,
\begin{equation}
m_N = m_N^0 + \gamma_N G_S \int\frac{d^3p}{(2\pi)^3}\frac{m_N}{E}
\left[ 1 - 2n_f(m_N;T) \right]\,,
\label{nnjlgap}
\end{equation}
where $m_N^0$ denotes explicit symmetry breaking mass and $G_S$ the strength
of scalar four-fermion interaction. Another parameter is a momentum cutoff
$\Lambda$ which regularizes the integral in Eq.~(\ref{nnjlgap}).
The parameters fixed to reproduce Eqs.~(\ref{binding}) and (\ref{incompress})
at zero temperature are given to be $\Lambda = 0.4$ GeV, 
$G_S = 1.677\,\mbox{GeV}\mbox{fm}^3$ and $m_N^0 = 41.3$ MeV~\cite{nnjl}. 
At $T=0$ in addition to the nuclear liquid-gas transition this model also
predicts the chiral crossover-type transition at $\rho \sim 3\,\rho_0$.
Here we check the prediction of this model at finite $T$.
Assuming a second-order chiral transition in the chiral limit, $m_N \to 0$
for $m_N^0 = 0$, the gap equation determines the critical temperature as
\begin{eqnarray}
&&
\frac{\pi^2}{12}T^2 + T^2 L_2[-e^{-\Lambda/T}]
{}- T\Lambda \ln\left( 1 + e^{-\Lambda/T}\right)
\nonumber\\
&&
= \frac{\Lambda^2}{8} - \frac{\pi^2}{2\gamma_N G_S}\,,
\end{eqnarray}
where $L_2[z] = \sum_{n=1}^{\infty} z^n/n^2$ is Euler's dilogarithm.
Substituting the parameter values one finds for the critical temperature
$T_c = 450$ MeV.
Numerical calculation with explicit breaking leads to a crossover
with slightly higher
temperature $T_c = 510$ MeV. Obviously this is too high to be accepted
as a reliable critical temperature compatible with the Lattice QCD 
predictions~\cite{latticeTc}. On the other hand, this $T_c$ is above the cutoff 
and therefore the model can not be used near the transition region.
Although larger $\Lambda$ can cure this drawback, another problem appears,
i.e. the bare nucleon mass becomes lower than three times
the bare quark mass~\cite{nnjl}.

In contrast, the parity doublet model leads to $T_c \sim 190$ MeV
at which the nucleon and its negative-parity partner remain massive
$m_{N+} \sim m_{N-} \sim m_0$. Therefore, this model satisfies the relevant 
phenomenological constraints not only on nuclear matter properties
but also on the chiral crossover at small chemical potential in a reasonable
way.

%%%%%%%%%%%%%%%%%%%%%%%%%%%%%%%%%%%%%%%%%%%%%%%%%%%%%
\section{Phase structure for large $N_c$}
\label{sec:nc}
%%%%%%%%%%%%%%%%%%%%%%%%%%%%%%%%%%%%%%%%%%%%%%%%%%%%

An interesting issue to address is how the phase structure is modified
by increasing the number of colors $N_c$. In large $N_c$ limit baryons arise as 
a soliton in the Skyrme model picture. In the current hadronic model 
we deal with the baryons as elementary particles, and thus there is no direct 
access to its dynamical description in the large $N_c$ limit, where microscopic
structure of the baryon is particularly important. Besides, it still remains 
uncertain at which scale a Skyrmion collapses (see e.g. Ref.~\cite{hkmp}
and references therein). Nevertheless, it is worthwhile to study a tendency
for various $N_c$, which may partly bridge a gap between the real and large 
$N_c$ QCD.

Following Ref.~\cite{witten} we adopt the $N_c$ dependences in the
model parameters as
\begin{eqnarray}
&&
m_{N\pm} \to (N_c/3)\,m_{N\pm}\,,
\quad
m_0 \to (N_c/3)\,m_0\,,
\nonumber\\
&&
a \to \sqrt{N_c/3}\,\,a\,,
\quad
b \to \sqrt{N_c/3}\,\,b\,,
\nonumber\\
&&
g_\omega \to \sqrt{N_c/3}\,\,g_\omega\,,
\quad
f_\pi \to \sqrt{N_c/3}\,\,f_\pi\,,
\nonumber\\
&&
\lambda \to (3/N_c)\,\lambda\,,
\end{eqnarray}
and the meson masses and the parameter $\bar{\mu}$ do not carry $N_c$
dependences. Inserting these scaling factors into the gap equations, one can 
perform thermodynamical calculations. In the following, without loss of generality,
we take zero temperature and consider the liquid-gas transition
and chiral symmetry restoration in cold nuclear matter.

\begin{figure}
\begin{center}
\includegraphics[width=8cm]{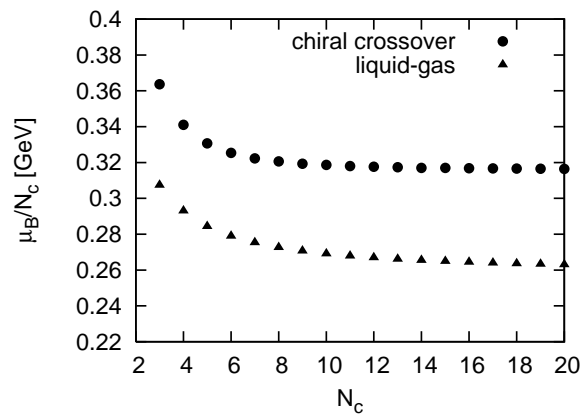}
\caption{
The (pseudo)critical chemical potentials for chiral
(circle) and liquid-gas (triangle) for various $N_c$.
The negative parity state was taken to be $N_-(1200)$.
}
\label{ncdep}
\end{center}
\end{figure}

As seen in section~\ref{sec:thermo}, how far the two transitions are
separated essentially relies on the mass of the negative parity state:
The liquid-gas transition from vacuum to nuclear matter ground state
happens roughly at $\mu_B \sim m_{N+}$. When its
parity partner is taken to be $N(1535)$, the initial mass gap between
the parity partners, which is generated by the spontaneous chiral 
symmetry breaking, is rather large $\sim 560$ MeV.
Compared with the case taking $N(1200)$, more matter compression is required 
in order to decrease the gap and consequently to restore
the chiral symmetry.
This mass gap between the parity partners scales linearly in $N_c$ and therefore
no coincidence of the liquid-gas and chiral phase transitions is generally
expected. In fact, the actual calculation in this model follows this 
consideration, as shown in Fig.~\ref{ncdep}.
%%%%%%%%%%%%%%%%%%%%%%%%%%%%%%%%%%%%%%%%%%%%%
%%%%%%%%%%%%%%%%%%%%%%%%%%%%%%%%%%%%%%%%%%%
With increasing $N_c$
the two phase transitions are systematically shifted toward higher 
chemical potentials. One sees the linear-$N_c$ behavior
in $\mu_B$ for around $N_c > 10$. 
On the other hand, if the effective coupling constants increase less
slowly with $N_c$ as discussed in Ref.~\cite{hkmp}, the two phase
transitions may coincide in the large $N_c$ limit~\cite{tm}.
More microscopic view of the baryons
at large $N_c$ might change this prediction, but this study is 
beyond the scope of this paper.

Deconfinement dynamics which is not dealt with in the current model
might also modify the above result, especially the phase boundaries
at higher temperatures. In fact, a pure NJL model for constituent quarks
predicts a chiral phase boundary of elliptic shape for any $N_c$. While,
NJL models with Polyakov loops (PNJL)~\cite{pnjl} describes a hierarchy in
quark and gluon sectors with respect to $N_c$ so that the deconfinement phase 
transition always happens at the critical value in a pure gluon system, 
$T_0 \sim 280$ MeV~\cite{mrs}.

%%%%%%%%%%%%%%%%%%%%%%%%%%%%%%%%%%%%%%%%%%%%%%%%%%%%%%%
\section{Conclusions}
\label{sec:conc}
%%%%%%%%%%%%%%%%%%%%%%%%%%%%%%%%%%%%%%%%%%%%%%%%%%%%%

We have investigated thermodynamics of nuclear matter in a parity doublet model
within the mean field approximation. The model describes the nuclear matter
ground state at zero temperature and a chiral crossover at zero chemical
potential at a reasonable temperature, which are the minimal 
requirements to describe the QCD thermodynamics. The first-order phase 
transitions appear only at low temperatures, below $T \sim 30$ MeV. 
Nevertheless, at higher temperature they still affect the order parameter 
which exhibits a substantial decrease near the liquid-gas {\it and} chiral 
transitions. If the chiral symmetry restoration is of first order, 
criticality around the end points of the two first-order phase transitions
will be the same due to the identical universality class~\cite{cp}.

The trace of the nuclear liquid-gas transition is
mostly overlapped with the meson-to-baryon ``transition'' characterizing
the ratio of the particle number densities, $\rho_M/\rho_B = 1$.
Its trajectory has only a weak parameter-dependence and the intersection
to the chiral crossover appears at $(T,\mu_B) \sim (150,450)$ MeV.
The chiral phase boundary at lower temperature, on the other hand, is
strongly affected by the negative parity state. The heavier parity partner
induces a larger baryon-rich domain in Nambu-Goldstone (NG) phase and thus
a strong enhancement in the baryon density is not necessarily accompanied by
the chiral symmetry restoration. The true parity partner of the nucleon needs
to be identified.
This requires to solve the model dynamically, since 
in NG phase the chiral symmetry does not dictate any non-trivial relations 
between the masses of parity partners and their couplings~\cite{jaffe}.

When increasing $N_c$ the two phase transitions are shifted to higher chemical
potentials and remain separated. This tendency is similar to the observation
in the PNJL model~\cite{mrs}: The threshold at which the baryon number density 
becomes finite and a first-order chiral phase transition do not coincide 
for any $N_c$ but are close with each other. 
This could indicate that the quarkyonic matter~\cite{quarkyonic} is realized 
in the chirally restored phase with confinement. However, this seems to
be unlikely, without introducing any gapless modes in a Wigner-Weyl phase,
since chiral anomalies in low-energy theories do not match with those in 
QCD~\cite{sasaki}. 
Further investigations along with the trace anomaly,
which is also responsible for appearance of a scale in QCD, may bridge the 
thermodynamics of the toy model used in this paper to the origin of $m_0$.

%%%%%%%%%%%%%%%%%%%%%%%%%%%%%%%%%%%%%%%%%%%%%%%%%%%%%%%%
\subsection*{Acknowledgments}
%%%%%%%%%%%%%%%%%%%%%%%%%%%%%%%%%%%%%%%%%%%%%%%%%%%%%%

We acknowledge stimulating discussions with M.~Lombardo,
K.~Redlich, L.~M.~Satarov, S.~Schramm, G.~Torrieri and W.~Weise.
This work has been partly supported by the Hessian LOEWE initiative 
through the Helmholtz International Center for FAIR (HIC for FAIR),
and by the grants NS-7235.2010.2 and RFBR 09-02-91331 (Russia).
C.S. thanks the organizers of the programme ``New Frontiers in QCD'' 
for their kind hospitality at Yukawa Institute for Theoretical Physics.

%%%%%%%%%%%%%%%%%%%%%%%%%%%%%%%%%%%%%%%%%%%%%%%%%%%%%%%%
%%%%%%%%%%%%%%%%%%%%%%%%%%%%%%%%%%%%%%%%%%%%%%%%%%%%%%%%%


\begin{thebibliography}{50}
\bibitem{qmproc}
C.~Sasaki,
  Nucl.\ Phys.\  A {\bf 830}, 649C (2009),
and references therein.

\bibitem{walecka}
  B.~D.~Serot and J.~D.~Walecka,
  Int.\ J.\ Mod.\ Phys.\  E {\bf 6}, 515 (1997).

\bibitem{other}
  J.~Boguta,
  Phys.\ Lett.\  B {\bf 120}, 34 (1983),
 I.~Mishustin, J.~Bondorf and M.~Rho,
  Nucl.\ Phys.\  A {\bf 555}, 215 (1993),
 G.~W.~Carter and P.~J.~Ellis,
  Nucl.\ Phys.\  A {\bf 628}, 325 (1998),
P.~Papazoglou, S.~Schramm, J.~Schaffner-Bielich, H.~Stoecker and W.~Greiner,
  Phys.\ Rev.\  C {\bf 57}, 2576 (1998),
P.~Papazoglou, D.~Zschiesche, S.~Schramm, J.~Schaffner-Bielich, H.~Stoecker and W.~Greiner,
  Phys.\ Rev.\  C {\bf 59}, 411 (1999).

\bibitem{nnjl}
  V.~Koch, T.~S.~Biro, J.~Kunz and U.~Mosel,
  Phys.\ Lett.\  B {\bf 185}, 1 (1987),
  M.~Buballa,
  Nucl.\ Phys.\  A {\bf 611}, 393 (1996),
  I.~N.~Mishustin, L.~M.~Satarov and W.~Greiner,
  Phys.\ Rept.\  {\bf 391}, 363 (2004).

\bibitem{bentz}
  W.~Bentz and A.~W.~Thomas,
  Nucl.\ Phys.\  A {\bf 696}, 138 (2001),
  W.~Bentz, T.~Horikawa, N.~Ishii and A.~W.~Thomas,
  Nucl.\ Phys.\  A {\bf 720}, 95 (2003).

\bibitem{dk}
  C.~E.~Detar and T.~Kunihiro,
  Phys.\ Rev.\  D {\bf 39}, 2805 (1989).

\bibitem{mirror}
  Y.~Nemoto, D.~Jido, M.~Oka and A.~Hosaka,
  Phys.\ Rev.\  D {\bf 57}, 4124 (1998),
  D.~Jido, Y.~Nemoto, M.~Oka and A.~Hosaka,
  Nucl.\ Phys.\  A {\bf 671}, 471 (2000),
  D.~Jido, T.~Hatsuda and T.~Kunihiro,
  Phys.\ Rev.\ Lett.\  {\bf 84}, 3252 (2000),
  D.~Jido, M.~Oka and A.~Hosaka,
  Prog.\ Theor.\ Phys.\  {\bf 106}, 873 (2001).

\bibitem{lsma1}
  S.~Gallas, F.~Giacosa and D.~H.~Rischke,
  arXiv:0907.5084 [hep-ph].

\bibitem{hp}
  T.~Hatsuda and M.~Prakash,
  Phys.\ Lett.\  B {\bf 224}, 11 (1989).

\bibitem{pdm}
  D.~Zschiesche, L.~Tolos, J.~Schaffner-Bielich and R.~D.~Pisarski,
  Phys.\ Rev.\  C {\bf 75}, 055202 (2007).

\bibitem{astro}
  V.~Dexheimer, S.~Schramm and D.~Zschiesche,
  Phys.\ Rev.\  C {\bf 77}, 025803 (2008),
  V.~Dexheimer, G.~Pagliara, L.~Tolos, J.~Schaffner-Bielich and S.~Schramm,
  Eur.\ Phys.\ J.\  A {\bf 38}, 105 (2008).

\bibitem{compress}
  J.~P.~Blaizot,
  Phys.\ Rept.\  {\bf 64}, 171 (1980),
  D.~Vretenar, T.~Niksic and P.~Ring,
  Phys.\ Rev.\  C {\bf 68}, 024310 (2003),
  D.~H.~Youngblood, Y.~W.~Lui, H.~L.~Clark, B.~John, Y.~Tokimoto and X.~Chen,
  Phys.\ Rev.\  C {\bf 69}, 034315 (2004).

\bibitem{qqbar}
  T.~D.~Cohen, R.~J.~Furnstahl and D.~K.~Griegel,
  Phys.\ Rev.\  C {\bf 45}, 1881 (1992).

\bibitem{sigma}
  J.~Gasser, H.~Leutwyler and M.~E.~Sainio,
  Phys.\ Lett.\  B {\bf 253}, 252 (1991).

\bibitem{chpt}
  N.~Kaiser, P.~de Homont and W.~Weise,
  Phys.\ Rev.\  C {\bf 77}, 025204 (2008).

\bibitem{triple}
  A.~Andronic {\it et al.},
  Nucl.\ Phys.\  A {\bf 837}, 65 (2010).

\bibitem{mrs}
  L.~McLerran, K.~Redlich and C.~Sasaki,
  Nucl.\ Phys.\  A {\bf 824}, 86 (2009).

\bibitem{quarkyonic}
  L.~McLerran and R.~D.~Pisarski,
  Nucl.\ Phys.\  A {\bf 796}, 83 (2007),
  Y.~Hidaka, L.~D.~McLerran and R.~D.~Pisarski,
  Nucl.\ Phys.\  A {\bf 808}, 117 (2008).

\bibitem{latticeTc}
  M.~Cheng {\it et al.},
  Phys.\ Rev.\  D {\bf 81}, 054510 (2010),
  S.~Borsanyi, Z.~Fodor, C.~Hoelbling, S.~D.~Katz, S.~Krieg, C.~Ratti and K.~K.~Szabo  [Wuppertal-Budapest Collaboration],
  arXiv:1005.3508 [hep-lat].

\bibitem{hkmp}
  Y.~Hidaka, T.~Kojo, L.~McLerran and R.~D.~Pisarski,
  arXiv:1004.2261 [hep-ph].

\bibitem{witten}
  E.~Witten,
  Nucl.\ Phys.\  B {\bf 160}, 57 (1979).

\bibitem{tm}
G.~Torrieri and I.~Mishustin, paper in preparation.

\bibitem{pnjl}
K.~Fukushima,
  Phys.\ Lett.\  B {\bf 591}, 277 (2004),
C.~Ratti, M.~A.~Thaler and W.~Weise,
  Phys.\ Rev.\  D {\bf 73}, 014019 (2006).

\bibitem{cp}
  A.~M.~Halasz, A.~D.~Jackson, R.~E.~Shrock, M.~A.~Stephanov 
  and J.~J.~M.~Verbaarschot,
  Phys.\ Rev.\  D {\bf 58}, 096007 (1998).

\bibitem{jaffe}
  R.~L.~Jaffe, D.~Pirjol and A.~Scardicchio,
  Phys.\ Rev.\ Lett.\  {\bf 96}, 121601 (2006);
  Phys.\ Rev.\  D {\bf 74}, 057901 (2006);
  Phys.\ Rept.\  {\bf 435}, 157 (2006).

\bibitem{sasaki}
  C.~Sasaki,
  arXiv:0910.4375 [hep-ph];
  arXiv:1004.5299 [hep-ph].


\end{thebibliography}
\end{document}